\documentclass[12pt,preprint]{aastex}

\begin{document}

\slugcomment{DRAFT - submitted to AJ }

\shorttitle{A $\delta$ Scuti distance to the LMC}
\shortauthors{McNamara et al.}

\title{A $\delta$ SCUTI DISTANCE TO THE LARGE MAGELLANIC 
CLOUD\altaffilmark{1}}

\author{D.~Harold~McNamara}
\affil{Department of Physics and Astronomy, Brigham Young University, N283 ESC,
Provo, UT 84602. 
email: harold\_mcnamara@byu.edu} 
\author{Gisella~Clementini}
\affil{INAF - Osservatorio Astronomico di Bologna, Via Ranzani 1, I-40127 
Bologna, Italy. 
email: gisella.clementini@oabo.inaf.it} 
\author{Marcella~Marconi}
\affil{INAF - Osservatorio Astronomico di Napoli, Via Moiariello 16, I-80131 
Napoli, Italy. email: marcella@na.astro.it} 
\altaffiltext{1}{Based on data collected at the European Southern Observatory,
proposal numbers 62.N-0802, and 66.A-0485} 

\begin{abstract}
We present results from a well studied $\delta$ Scuti star discovered in the 
Large Magellanic Cloud (LMC). The absolute magnitude of 
the variable was determined from the Period-Luminosity ($P-L$) relation for 
Galactic $\delta$ Scuti stars and from the 
theoretical modeling of the observed $B,V,I$ light curves with nonlinear 
pulsation models. The two methods give distance moduli for the
LMC of 18.46$\pm$ 0.19 and 18.48$\pm$ 0.15, respectively, for a consistent 
value of the stellar reddening of $E(B-V)$=0.08$\pm$0.02.

We have also analyzed 
24 $\delta$ Scuti 
candidates discovered in the OGLE II survey of the LMC, 
and 7 variables identified in the open cluster LW 55 and in the galaxy disk by Kaluzny et al. (2003, 
2006). We find that the LMC  $\delta$ Scuti stars define a $P-L$ relation whose slope
is very similar to that defined by the Galactic $\delta$ Scuti variables,  
and yield a distance modulus for the LMC of  18.50$\pm$0.22 mag.  
We compare  the results obtained from the $\delta$ Scuti variables with those derived 
from the LMC RR Lyrae stars and Cepheids.  
The corresponding distance moduli are:  
$\delta$ Scuti stars 18.48 $\pm$ 0.02 mag (standard deviation of the weighted average
of the three above solutions); RR Lyrae stars 18.49$\pm$0.06 mag; and 
Cepheids 18.53$\pm$0.02 mag.  We have 
assumed an average color excess of $E(B-V)$ = 0.08 mag $\pm$ 0.02  
% and a metallicity correction of $-0.19 \times$ [Fe/H] 
for 
both $\delta$ Scuti stars and Cepheids.
Within the observational uncertainties, the three
groups of pulsating stars yield very similar distance moduli.  
These moduli are all consistent 
with the ``long" astronomical distance scale for the Large Magellanic Cloud.
\end{abstract}

\keywords{stars: oscillations $-$ (stars: variables:)$\delta$ Scuti $-$ stars: variables:
other $-$ galaxies: distances and redshifts $-$ galaxies: Magellanic Clouds   
$-$ (cosmology:) distance scale}

\section{INTRODUCTION}

The $\delta $ Scuti variables are late A and early F-type stars that 
populate the instability strip near or slightly above the zero-age 
main-sequence, from luminosities $M_{V }\sim $ 3.0 mag to $M_{V} \quad \sim $ 0.5 
mag. Both radial and non-radial pulsation modes have been detected in these 
stars. The large-amplitude ( $A_{V } >$ 0.25 mag)  variables  have been referred to in the literature
as dwarf Cepheids, HADS (high amplitude $\delta $ Scuti) and SX Phe stars 
(Population II dwarf Cepheids) if they are metal poor. 
The large-amplitude variables with asymmetrical 
light curves are radial pulsators pulsating usually in the fundamental mode,  
but some exhibit pulsation in both the fundamental and first harmonic. The 
variables found in globular clusters tend to be fundamental mode pulsators if 
they have large asymmetric light curves, while the stars with large 
symmetric light curves or small amplitude variables, $A_{V }<$ 0.1 mag, 
tend to be pulsating in higher modes (usually the first harmonic; McNamara, 
2000). An excellent review of the properties of these stars can be found in 
Breger (2000).
Since these variables obey a Period-Luminosity ($P-L$) relation they can be 
utilized as standard candles to find distances.

Given their intrinsic faintness and the short time scale of their 
variations (typical periods range from 1 to 5 hours), the known samples of 
$\delta$ Scuti stars almost exclusively belong to the Milky Way where they 
are found in the field and in both globular and open clusters.
$\delta$ Scuti or dwarf Cepheids found in globular clusters are Population II dwarf Cepheids 
(SX Phoenicis). Nemec, Linnell Nemec, \& Lutz (1994) present a comprehensive review of 
the Population II
dwarf Cepheids (SX Phoenicis) known at the time. They all were located in
 Galactic globular clusters (NGC~5053, $\omega$ Cen, NGC~5466, M71). 
 Many more have been discovered since (see the catalog of 
 Rodr\'{\i}guez, L\'opez-Gonz\'alez 2000).
Many $\delta$ Scuti stars have been found in the field and in open clusters 
(Rodr\'{\i}guez, L\'opez-Gonz\'alez, \& L\'opez de Coca 2000). 

Since their detection requires time-consuming observations reaching
 about 2-3 magnitudes below the horizontal branch (HB) of the old population,
after the pioneering studies in Carina
by Mateo, Hurley-Keller, \& Nemec (1998) and Poretti (1999) who reanalyzed the Carina 
dwarf Cepheids ($\delta$ Scuti stars) discovered by Mateo et al., so far only few further of 
these
variables have been identified in systems external to
the Milky Way. Namely, a $\delta$ Scuti star 
with $P\sim$0.11 d was discovered in the LMC by Clementini 
et al. (2003, hereinafter C03) and $B,V,I$ photometry was published by Di Fabrizio et 
al. (2005, hereinafter
DF05), 6 $\delta$ Scuti candidates were found in the LMC open cluster LW 55 by 
Kaluzny \& Rucinski (2003), one in the LMC disk by Kaluzny, 
Mochnacki \& Rucinski (2006), and further 24 candidates 
were identified in a list of short-period variables found in the OGLE II
RR Lyrae catalog of the LMC (Soszy\'nski et al. 2003).
Finally, about 90 SX Phoenicis variables 
have 
been found in the Fornax 
dwarf spheroidal galaxy by Clementini et al. (2004) and Poretti et al. (2006a,b).

In this paper we present results from the analysis of the $\delta$ Scuti star discovered 
in the LMC by C03 and DF05, and compare it with the other $\delta$ Scuti candidates detected
in 
the LMC.
In Section~2 we describe the observational database on which the present analysis
is based. In Section~3 we discuss the $P-L$ relation of the LMC 
$\delta$ Scuti stars, the absolute magnitude we derive for both 
the program star and the 
other LMC $\delta$ Scuti candidates, and
the corresponding distance modulus for the LMC. Then we use the $\delta$ Scuti
stars and the Cepheids in the LMC to explore the possibility that the two groups
of stars obey the same $P-L$ relation. Section~4 is devoted to the 
determination of the absolute magnitude of
our target star from the theoretical reproduction of its light curves, with 
nonlinear convective pulsation models (Bono et al. 1997, Bono, Castellani, 
Marconi 2000). 
The comparison between the LMC distance
 moduli obtained by  
% two independent techniques 
the $\delta$ Scuti $P-L$ relation and by the model fitting technique  
is presented 
 in Section~5, along with a discussion of results 
obtained from other pulsating stars in the LMC, namely RR Lyrae stars 
and Cepheids.

\section{THE OBSERVATIONAL DATABASE}

The analysis presented in this paper is based on $B,V,I$ CCD observations 
in the Johnson-Cousins photometric system of a high
amplitude $\delta$ Scuti star in the LMC obtained 
with the 1.5 m ESO telescope, La Silla, Chile, in
1999 and 2001 by C03 and DF05. 
The $\delta$ Scuti variable, star \#28114 in DF05 catalog
($\alpha_{2000}$=05 22 35.5 $\delta_{2000}$=$-$70 28 15.5),
was detected in 
Field A of C03 and DF05. This is 
the field closer to the LMC bar, for which C03
derived an average reddening $E(B-V)$=0.116$\pm$0.017 mag from the properties
of the RR Lyrae stars. The observational characteristics of 
the program star are summarized in Table~\ref{t:table1}, and  
the light curves are shown 
in Figure~1. 
They have 69, 40, and 11 phase points in $V$, $B$, and $I$, 
with corresponding photometric accuracies of  0.09, 0.09 and 0.06 mag, 
respectively\footnote{These uncertainties include the contribution of the
internal photometric accuracy, computed as the average of the 
residuals from the Fourier best fitting models of the light curves (
0.087, 0.077, and 0.042 mag in $V$, $B$ and $I$, respectively), 
the uncertainties of the photometric calibration (0.018, 0.032 and 0.025 mag), and 
of the aperture corrections (0.018, 032 and 0.030 mag), see C03 and DF05 for details.}. 
Figure~2 shows the position of the star in the color magnitude
diagram of the LMC Field A (see DF05 and C03).

\section{THE $\delta$ SCUTI ABSOLUTE MAGNITUDE FROM THE PERIOD-LUMINOSITY
RELATION}

The most recent $P-L$ relation of $\delta$ Scuti stars is that given by McNamara
et al. (2004), that was derived from 48 Galactic $\delta$ Scuti stars with
accurate Hipparcos parallaxes (errors in the parallaxes $\leq$ 10\% of
the parallax).
We will utilize the first equation in the paper that is valid for [Fe/H]$\geq -1.5$, namely
		
		$$M_{V} = -2.90 \times \log P-0.190 \times {\rm [Fe/H]}-1.27,\,\,\,\,
		\sigma=0.16, \,\,\,\,{\rm (1)}$$
where $P$ is the fundamental mode period expressed in days. 
The relation has a dispersion of 0.16 mag, that 
we assume as the error in the $M_{V}$ estimates obtained through eq. (1).
We have introduced a small change in the zero point ($-$1.26 to $-$1.27) based on  
current studies. 
 Recently, Buzasi et al.(2005) have found Altair to be a $\delta$ Scuti star.  
Equation (1) predicts for the star an absolute magnitude of $M_{V}$ = 2.204 
($\log P = -1.1978$ and 
[Fe/H] = 0.00), which compares very well with $M_{V}$ = 2.208, derived from the star's 
very accurate parallax.

Although a paper giving more details regarding equation (1) is in preparation by one 
of us (D.H.M.), a few details are given here. This equation is superior to other equations 
derived in the past 
because it introduces a metallicity term.  $P-L$ relations derived previously without the 
metallicity term have much steeper negative slopes (see, for example, McNamara 1997  
and Petersen \& Christian-Dalsgaard 1999).  The steeper slope can be attributed to 
the fact that the metal-poor variables are fainter than the metal-rich ones and are 
found primarily at short periods.
Lutz-Kelker (1973) corrections were applied to the magnitudes calculated from the parallaxes. 
Figure~3 exhibits the data and the fit given by equation (1)
with $[Fe/H]=0.00$. 
The open square in Figure~3 is Altair (Buzasi et al. 2005), while the cross is the
metal-poor variable SX Phe ([Fe/H] $\sim -1.5$, McNamara 1997).
The mean error of the $M_V$ values due to parallax errors is 
$\pm$ 0.13 mag, indicating that the cosmic scatter in the relation is $\sim$0.09 mag
since the dispersion is $\pm$0.16 mag. The cosmic scatter is a consequence of the finite
width of the instability strip.
%
%Actually, most of the dispersion is due to errors in the parallax, as the cosmic scatter in the
%relation is $\sim$0.08 mag to 0.10 mag. Thus the error of $\pm$0.16 mag is likely a
%conservative estimate, perhaps $\pm$0.10 mag being a more realistic value.
%
%
%{\bf Although a paper giving more details regarding equation (1) is in preparation by one 
%of us (D.H.M.), a few details are given here.
%
%
We also show in Figure~4 the fit of equation (1) to Cepheids plus $\delta$ Scuti stars. 
The Cepheid $M_V$ values are from 
Kervella et al. (2004) interferometric angular diameter measurements, and Hoyle et al. 
(2003) Cepheids in open clusters (SV Vul excluded). 
Other Cepheid data give similar results.

The first to propose $\delta$ Scuti stars as ``small Cepheids'',  obeying the same 
period-luminosity law, was Fernie (1992).  
Utilizing $ubvy$ data to calculate the $M_V$ values of the $\delta$ Scuti stars and combining 
with Cepheid $M_V$ values he found 
$$M_V=-2.902( \pm 0.030)\log{P}-1.203( \pm 0.029),\,\,\,\,
		\sigma=0.26, \,\,\,\,{\rm (2)}$$
Laney, Joner \& Schwendiman (2002) used Baade-Wesselink radii of both $\delta$ Scuti stars and 
Cepheids to derive 
the corresponding $M_V$ values.
They found
$$M_V=-2.92( \pm 0.04)\log{P}-1.29( \pm0.04),\,\,\,\,
		\sigma=0.234, \,\,\,\,{\rm (3)}$$
All three equations ($[Fe/H]$=0.00 in eq. (1)) have similar slopes, but somewhat different 
zero points. Equations (1) and (3) predict 
exactly the same $M_V$ values over the $\log{P}$ interval -1.25$<{\log{P}}<{-0.75}$ and differ 
by only 0.01 mag outside this range where only a few $\delta$ Scuti stars 
are found. This is a remarkable agreement when  one considers that 
the two equations were derived by independent methods.
The slope  $-2.90$ of equation (1) differs from the one ($-3.14$) recently derived by Tammann 
et al. (2003) for  Cepheids, but agrees quite well with the value ($-2.87$) found by Laney 
\& Stobie (1994) for the same class of variables and with the theoretical slope 
($-2.905$) by Alibert et al. (1999) based on Cepheid linear nonadiabatic models.
Only equation (1) introduces a $[Fe/H]$ term that is valid only for the $\delta$ Scuti stars.
Stellar evolution and pulsation models (Petersen \& Christensen-Dalsgaard 1999; 
Santolamazza et al. 2001; Petersen \& Freehammer 2002) indicate that the
absolute magnitude of the $\delta$ Scuti stars depends on chemical composition.
We find that $\delta{M_V}/{\delta{[Fe/H]}}$ is $-$0.19
for the interval 0.0 $< [Fe/H] < -$1.4 and is $-0.089$ for the interval $-1.6 < [Fe/H] < -$2.5.
The inflection point occurs at $[Fe/H]=-1.5$. This is the reason why two separate 
equations are given 
in McNamara et al. (2004) for calculating the $M_V$ values of $\delta$ Scuti stars.
Templeton (2000) assumed that $\delta{M_V}/{\delta{[Fe/H]}}$ was linear over the whole range 
of $[Fe/H]$ values and found the coefficient to be -0.122, close to the mean of the two 
coefficients above ($-0.140$). Petersen \& Christensen-Dalsgaard (1999) found from an 
analysis of 22 $\delta$ Scuti stars for which stellar models had been used to infer
$M_V$, that $\delta{M_V}/\delta{[Fe/H]}=-0.177$. This value refers primarily to stars with 
$[Fe/H]>-1.4$, since only 2 stars had  $[Fe/H]$ values $< -1.5$ and compares favorably 
with our $-0.190$. Their population I models, on the other hand, yelded $-0.271$ over the 
$[Fe/H]$ interval 0.00 to $-0.85$.
Two papers deal directly with Population II stars. Santolamazza et al. 
(2001) found that $\delta{M_{bol}}/\delta{[Fe/H]}=-0.07\pm0.01$ over the 
interval $[Fe/H]=-1.5$ to $-2.5$ and Petersen \& Freehammer (2002) found  
$\delta{M_{bol}}/\delta{[Fe/H]}=-0.114$ over the interval $-2.5 \le [Fe/H] \le -0.85$. 
These values compare favorably with $\delta{M_V}/\delta{[Fe/H]}=-0.089$, the value we found 
for $[Fe/H]<-1.5$. The metal-poor 
%($[Fe/H] \sim -1.5$) 
variable star SX Phe 
%is marked by a 
(cross   
in Fig. 3) falls well below the Population I variables, as predicted by model calculations.

The LMC variable discovered by C03 and DF05 (\#28114) is a large amplitude 
variable with very symmetrical light 
curves as is evident in Figure~1. This suggests the star is pulsating 
in the first harmonic or a higher mode.  
 In what 
follows we assume that star is a first harmonic pulsator, accordingly the star's period 
in Table~\ref{t:table1} has been named $P_1$.  Actually, only as a first 
harmonic pulsator can we derive a reasonable distance to the LMC from the 
$P-L$ relation, and a good theoretical fit of the star's light curves (see Section 4).  
We will show later 
that when the period is fundamentalized the star fits the $P-L$ relation 
defined by the other $\delta$ Scuti candidates in the LMC.
The $P_{1}$ value, column 3 of Table~\ref{t:table1}, is fundamentalized by adopting 
the ratio of the first harmonic to the fundamental period ($P$), $P_{1}/P = $0.773. 
This ratio has been well established from double-mode 
pulsators in the Galaxy (Petersen \& Christensen-Dalsgaard, 1996). The 
corresponding fundamental period is $\log P = -0.836$.
We do not have a direct estimate of the metallicity of star \# 28114.
Since in the case of the Galaxy the stars with $\log{P}$ longer than -1.1 are restricted 
pretty well to $[Fe/H]$ near 0.0, we assume that the same is true in the case of 
the LMC stars with $\log{P}$ values greater than -1.1. In particular, 
we assume that star \# 28114 
%the LMC $\delta $ 
%Scuti star 
is typical of the intermediate age population in the LMC where 
[Fe/H] = $-$0.40 
(Russell \& Dopita 1990). This is also the average metallicity of the LMC Cepheids 
(Luck et al. 1998). 
 The M$_{V}$ value of star \# 28114 
following from equation (1) is then M$_{V} =$1.23 $\pm 0.16$ mag.  
A change of $\pm$0.1 dex in metallicity would cause a change of
$\pm$0.02 mag in the $M_V$ value of star \# 28114 derived through eq. (1).

 According to C03 
the average color excess of RR Lyrae stars in the LMC bar in the direction 
of the variable is $0.116 \pm 0.017$, leading to an absorption of 
$A_{V}$ = 0.36 mag ($A_{V}= 3.1 E (B-V)$ assumed). The apparent magnitude of the 
$\delta $ Scuti star is $\left< V \right>$ = 19.94 $\pm$ 0.09 mag (Table~\ref{t:table1}), 
that combined with the absolute magnitude of $M_{V}$ = 1.23 $\pm$0.16 mag, and with the 
visual absorption of A$_{V}$ = 0.36 $\pm$ 0.05 mag, leads 
to a true distance modulus for the LMC of $\left< V_0 \right> - M_{V}$ = 18.35 $\pm$ 0.19 mag,
(where the error is the sum in quadrature of the uncertainties in the photometry, reddening and
absolute magnitude).

An alternative approach is to employ directly the observed $\left< B-V \right>$ value 
of the $\delta$ Scuti variable to infer its color excess. 
Galactic $\delta $ Scuti stars indicate that $\left< b-y \right>_{0}$=0.185 $\pm$ 0.015 
mag at the period of \#28114 (McNamara 1997).
If we adopt $\left< b-y \right>_{0}$=0.17 (we allow for a decrease
in $\left< b-y \right>_{0}$ for a decrease in metallicity) and assume $m1 \simeq$  0.145 mag  
as appropriate for [Fe/H] = $-$0.4, then the equation: 

		$$B-V = 1.505 \times (b-y)+ 0.576 \times m1 - 0.083,\,\,\,\,{\rm (4)}$$

given by Cousins and Caldwell (1985) to convert 
to $B-V$, yields $(B-V)_0$ = 0.256 mag.
This value of $(B-V)_0$ when combined with the 
observed $B-V$ = 0.332 mag suggests a color excess of $E(B-V)$ = 0.076 $\pm 0.020$ mag 
for star 
\# 28114. This color excess agrees well with 
the reddening maps of Schlegel, et al. (1998) which give a mean foreground reddening of the LMC 
of $E(B-V)$ = 0.075 $\pm 0.02$ mag. Schlegel, et al. value is the average 
of $E(B-V)$ measurements in 
annular rings about the LMC center. 
Since there is little change from ring to ring it represents a very meaningful value
for the foreground reddening of the LMC. On the other hand, 
the Burnstein \& Heiles (1982) maps indicate that the LMC is located in the color excess 
contours of 0.06 and 0.09 mag, thus supporting 
the Schlegel et al. mean value of 0.075.
The observational evidence thus seems to indicate that the color excess of this variable 
cannot be as large as 
0.116 mag. In fact, if $E(B-V)$=0.116, the  $(B-V)_0$ color of star \# 28114 would be too 
blue (0.216 mag). Even extremely metal-poor variables have redder $(B-V)_0$ values, of the
order of 0.25 mag.
Further, we anticipate that also the ``pulsational'' reddening derived
from the light curve fitting (see Section 4) 
is smaller than C03 average 
value for the area,
and very close to the reddening inferred from the star intrinsic color
($E(B-V)_{puls.}$=0.07$\pm$ 0.14).
Given the patchy nature of the reddening in the LMC it is possible that our target star 
has lower than average reddening. Another possibility is that the star
is a foreground object subject to a smaller extinction. Finally,
a third possibility is that the spatial distribution along the line-of-sight of the old stars 
(namely the RR Lyrae variables) from which C03 reddening was derived is different from 
that of much younger objects as the $\delta$ Scuti stars or the Cepheids. 
For instance, an average value $E(B-V)$=0.07 is obtained by considering individual 
Cepheids in the LMC (see Caldwell \& Coulson 1986, Gieren, Fouqu\'e \& Gomez 1998, 
and the discussion in C03).  
Based on the above considerations we adopt for star \#28114 $E(B-V)$ = 0.08 $\pm$ 0.02 mag.  
This leads to a distance modulus of 18.46 $\pm$ 0.19 mag, where again the error 
is the sum in quadrature of the uncertainties in the reddening, absolute magnitude and 
photometry.

\subsection{THE PERIOD-LUMINOSITY RELATION OF THE LMC $\delta$ SCUTI STARS}
We will now use star \#28114 along with 
the 
other $\delta$ Scuti candidates detected in the LMC in an attempt to define the $P-L$ relation of
the LMC $\delta$ Scuti stars. We discuss first 
 the list of short-period variables discovered in 
the OGLE II survey of the LMC. These stars are classified as ``other pulsating variables''
in Soszy\'nski et al. (2003) catalog of the LMC RR Lyrae stars. They  
are given in a table (other\_ogle.tab) that does not appear directly in the paper
and  must be downloaded from the OGLE Internet archive for the LMC RR Lyrae
stars\footnote{Available at ftp://sirius.astrouw.edu.pl/ogle/ogle2/var\_stars/lmc/rrlyr/}. 
The table lists 37 variables\footnote{Star OGLE54500.13 appears twice in the table 
namely as LMC\_SC19 star number 171019, and LMC\_SC20 star number 37093}
with periods shorter than $P \lesssim$0.5 d and generally 
around 0.2 d. The light curves of these 37 stars 
are generally poor, being some of the objects at the magnitude limit of OGLE II
photometry.
The authors provide 
intensity mean $\left< I \right>$, $\left< V\right>$, $\left< B \right>$ magnitudes and 
corresponding de-reddened $\left< I \right>_0$, $\left< V\right>_0$, $\left< B \right>_0$ 
values. 
The de-reddened magnitudes were computed assuming for each variable
the reddening of the field where the star is located \footnote{Individual reddenings are
given in Table~\ref{t:table1} of Soszy\'nski et al. (2003). They are the 
same as in Udalski et al. (1999a).}, 
and the extinction law $A_V$=3.24$E(B-V)$ (Soszy\'nski et al. 2003). 
In addition the periods and, where appropriate (i.e. for stars suspected to be double-mode 
pulsators), also the secondary periods are given. 

The mean $\left< V\right>$ magnitudes of these 37 stars are plotted against $\log P$ in 
Figure~5.
Single-mode pulsators are shown as filled circles, double-mode pulsators are shown as
filled triangles, stars vibrating in two periods 
in which the period ratio is $P_1/P \sim$ 0.76 are plotted as open squares.
For all double-mode pulsators the shortest periods have been used
in the plot. 
Figure~5 shows a very large scatter.
In an attempt to select the most likely $\delta$ Scuti stars we cleaned 
the sample eliminating objects in Figure~5 having P$\geq$0.22 d\footnote{The period of 
0.22 d was chosen because at longer periods RR Lyrae stars (first overtone pulsators, in 
particular), and $\delta$ Scuti stars have similar absolute magnitude. With this cut off
we have reduced the possibility of getting RR Lyrae stars in our sample} and/or brighter than $V\sim$19,
being the average luminosity of the RR Lyrae stars in the OGLE II survey:   
$\left< V\right> \sim$ 19.3 (Soszy\'nski et al. 2003). 
We also eliminated the faintest star in the list: star 324548 (OGLE52417.16-692429.3) 
since its magnitudes lead to an unrealistic $B-V$ of 0.02 mag.
The sample cleaned according to the above criteria includes 24 stars, they are
listed in the upper portion of Table~\ref{t:table2}. Column 5 of the table gives the 
OGLE II mean $\left< V\right>$ magnitudes 
of these stars. DF05 show that there are systematic differences between OGLE and
DF05 photometries based on the comparison of a large number of stars (about 6000) in common
in the magnitude range from $V \sim 16$ to $V \sim 20.5$. 
Reasons for these differences are widely discussed in Section 4.3 of DF05 paper.
DF05 magnitudes are on average systematically slightly fainter 
than OGLE's photometry for the same stars
because 
DF05 photometry is able to reach about 1.5 mag fainter and to resolve about 40\% more stars
stars than OGLE II in the same area, thanks to the higher performances 
to measure faint stars in crowded fields of the reduction 
procedures adopted by DF05 (DAOPHOT/ALLFRAME) compared to OGLE's (DoPhot). 
%DF05 photometry is on average slightly fainter 
%than OGLE's photometry for the same stars, since it is known that DoPhot gives
%systematically brighter magnitudes for faint stars in crowded regions than DAOPHOT/ALLFRAME,
%and since DF05 resolved many more faint stars than OGLE II in the area common 
At the luminosity level of the LMC 
$\delta$ Scuti stars DF05 photometry is about 0.06 mag fainter than OGLE's.
This difference decreases to about 0.04 for $V$ in the range from 15 to 18 mag 
(see Table 15 in DF05).
%Reasons for this difference are widely discussed in DF05 which the
%interested reader is referred to for details
%{\bf (Gisella, aggiungi qui dei pezzi presi dal paper DF05)
%DF05 photometry is found to be systematically fainter than OGLE II photometry for the same
%stars as expected since the DoPhot used by the OGLE II team is reported to give 
%systematically brighter magnitudes
%for faint stars in crowded in field than  
Since DF05 $V$ photometry is generally more accurate and deeper than OGLE's (see 
upper panels of Figure 25
in DF05) we have   
adopted DF05 photometric zero point, and have thus transformed 
OGLE II $\left< V\right>$ mean magnitudes using the transformation relations
provided by DF05. Corrected $\left< V_c\right>$ values are given in Column 7 
of Table~\ref{t:table2}.

In order to identify the pulsation mode of the OGLE stars in
Table~\ref{t:table2} it would be desirable 
to have precise light curves with a full frequency analysis available.  Unfortunately, 
such is not the case here.  We instead, have been forced to try 
and identify the pulsation modes.  We have used the following criteria:  
(1) stars with two frequencies and period ratios of P1/P0 $\sim$ 0.76 - 0.775 are 
pulsating in the fundamental and first overtone modes, (2) the greatest concentration 
of data points in a $\left< V\right>$ - $\log P$ plot are likely the fundamental pulsators, 
(3) the 
stars falling near the lower envelope in the plot must be pulsating in the fundamental 
mode,  (4) the stars should exhibit a $P - L$ relation with an expected cosmic 
scatter of $\sigma$ $\sim$ 0.09 magnitudes, (5) first and second overtone pulsators 
should lie $\sim$ 0.3 mag and 0.6 mag above the fundamental pulsators in the $\left< V\right>$ - $\log P$
 plot, respectively.  On the basis of these criteria we have assigned the variable's pulsation modes 
(see Column 9 of Table~\ref{t:table2}). They are subject to improvement when
  more precise data become available.

In Figure~6 we plot the corrected mean $\left< V_c\right>$ magnitudes of OGLE II
cleaned sample using open circles for stars likely 
pulsating in the first or higher overtones and filled circles for stars
pulsating in the fundamental mode. 
The group of six stars 
at 19.4$ <\left< V_c\right> <$ 19.7 mag and $\log P \sim -0.8$, which are likely 
pulsating in the first overtone, were fundamentalized assuming
$P_{1}/P = $0.773 and are also plotted as X's.
The four stars 
at 19.0$ <\left< V_c\right> <$ 19.4 mag and $\log P \sim -0.77$, are likely 
pulsating in the second overtone. Since 
$P_{1}/P = $0.773 and $P_{2}/P_{1} = $ 0.805 (see Figure 5 of Jurcsik et al. 2006) 
they were fundamentalized assuming $P_{2}/P = $0.6223 and are also plotted as
asterisks. The open triangle represents the position of the \#28114 discussed 
previously, according to its fundamentalized period. 
The best fit regression line directly computed on the fundamental/fundamentalized stars is shown by the
solid line that is described by: 

$$\left< V_c\right> = -2.577 (\pm 0.314) \times \log P + 17.814 (\pm 0.219), \,\,\,\, {\rm (5)}$$
with a rms scatter of 0.125 mag.

In Table~\ref{t:table3} we list periods, and mean magnitudes and colors for the 
$\delta$ Scuti stars identified in the LMC by Kaluzny \& Rucinski (2003) and Kaluzny et al.
(2006). For the LW 55 $\delta$ Scuti stars periods and pulsation modes 
(see columns 4 and 9 of Table~\ref{t:table3}) were estimated by
one of us (HMcN) from the charts given in Kaluzny \& Rucinski (2003) paper. 
Although, the LW55 stars appear to be multimodal, the light data do exhibit alternating light 
maxima and minima from which the period of the dominant mode can be inferred.
The error in the $\log P$ values of these stars is the order of $\pm$ 0.02.
So these periods are definitely approximate, still we think they are 
sufficiently accurate to use in finding slopes and intercepts from linear regressions.
The one star in Kaluzny et al. (2006) paper has a superb light curve and a very accurate period
derived by the authors. It exhibits a short period and a rather large
residual, in a linear regression solution (equation 7). We expect the star may have a much lower metallicity than the other objects  
discussed in this paper, similar to what happens in the Galaxy, where the short period 
stars typically are very metal poor (see Fig.1, McNamara 1997).
The $\left< V \right>$ and $\left< B-V\right>$ values in columns 5 and 6 are mean magnitudes
rather than intensity mean magnitudes, taken from Kaluzny et al. (2003, 2006). Since 
the amplitudes are very small this should not add any serious error. 
The $\left< B-V\right>$ values compare very favorably with the $\left< B-V\right>$ value of 
star \#28114. By adopting
$(B-V)_0$=0.256 it follows that the average color excess of the 7 stars is $E(B-V)$=0.085 mag
($\sigma$=0.025 mag).
This is in very good agreement with the value we have adopted through the paper: $E(B-V)$=0.08.

In Figure~7 we plot the $\left< V_c\right>$ magnitude of the 24 
$\delta$ Scuti candidates in OGLE II (filled circles), along with the $\left< V \right>$ 
magnitude of the
7 $\delta$ Scuti stars in Kaluzny et al. (2003, 2006; open circles), and with star \#28114 
(open triangle). Fundamental and fundamentalized periods are used for all of them.
Since we have no stars in common with Kaluzny et al. to check any systematics in the 
magnitude scales, 
we are forced to assume that the Kaluzny et al. magnitudes are on the same scale. 
The majority of the LMC $\delta$ Scuti discussed so far are located in the direction of 
the LMC bar.
The Kaluzny et al. (2003, 2006) variables are an exception, since they are located in the 
SE quadrant of the
LMC. We found it necessary to correct stars V1-V8 (LW55 Table~3) by $-0.10$ mag and star 
V2 (Disk Table~3)
by $-0.06$ mag to place them at the same distance as the bar variables.
Tilt-corrected magnitudes for these stars are provided in Column 6 of Table~3.
%......  E POI VANNO RICALCOLATE TUTTE LE REGRESSIONI DA QUI IN POI, TEMO!!})
%We would not expect any systematic differences to be to large because XXXX hanno usato anche
%loro ALLFRAME??? CHECK}.
The dashed line is
the $P-L$ relation where we force the slope to be $-2.90$ and fitting 
the 
% of the Galactic $\delta$ Scuti stars with  
%[Fe/H]$\leq -1.5$ (equation (1)) computed at [Fe/H]=$-0.40$ and shifted to 
%fit the 
fundamental
mode (and fundamentalized) LMC $\delta$ Scuti stars, its equation is:

$$\left< V\right> = -2.90 \times \log P + 17.574, \,\,\,\, {\rm (6)}$$
		
%{\bf This relation fits reasonably (????) 
%very
%well the 32 fundamental mode $\delta$ Scuti stars shown in Figure~7.}
The best fit regression line directly computed on the stars is shown by the
solid line, its equation is: 

$$\left< V_c\right> = -2.635 (\pm 0.156) \times \log P + 17.770 (\pm 0.121), \,\,\,\, {\rm (7)}$$
with a rms scatter of 0.134 mag.
 
%The two $P-L$ relations are almost indistinguishable.
The scatter around the $P-L$ relations in Figure~7 might be partially due to 
differential reddening affecting the various stars.
For a more rigorous derivation of both slope and zero point of the $P-L$ relation, and
to possibly reduce its scatter we have corrected for these effects.
It is well known that the reddening within the LMC varies from one region to the other,
thus stressing the need for ``local" estimates of the color excess. Several different 
way to measure the LMC reddening 
are found in the literature. A detailed discussion and comparison of different estimates 
can be found in Section 4 of
C03 paper. C03 color excess was derived by comparing the colors of the edges of the 
instability strip defined by
RR Lyrae stars in the field where our $\delta$ Scuti target is located,  to 
those of variables in globular clusters of known
$E(B-V)$. Estimates of reddening in the same area were obtained by OGLE-II using differences 
of the observed 
$I$-band magnitude
of red clump stars to map the color excess. OGLE-II reddening is on average about 0.03 larger 
than derived 
by C03 from the 
RR Lyrae stars in the same area. 
C03 reddening scale is preferred here because the adoption of OGLE-II reddening scale 
would result 
in too blue colors for the LMC RR Lyrae variables and would enhance the discrepancy for the 
$\delta$ Scuti star
too, for which we find that even C03 reddening might be an overstimate (see discussion in 
Section~3).
%Their reddening
%scale, is 0.03 mag lower on average than OGLE color 
%excess in the same area. 
%C03 reddening scale is preferred here because it was estimated from stars in the same 
%area where our $\delta$ Scuti target is located and . 
%%%\footnote{As widely discussed in C03, OGLE II reddening
%%%appears to be about 0.03 larger than derived by C03 from RR Lyrae stars in the same area.
%%%For the $\delta$ Scuti stars and Cepheids the difference could even be larger since  
%%%OGLE II extinction was determined from stars older than the Cepheids and the 
%%%$\delta$ Scuti stars, namely from old red clump stars. In fact, as suggested by Udalski et al. (1999a),
%%% it is possible that the
%%%spatial distribution of the red clump stars along the line-of-sight  is different from that 
%%%of much younger objects like Cepheids and $\delta$ Scuti stars.}.
We have thus 
corrected downward by
0.028 mag the individual reddenings 
of the OGLE II 
$\delta$ Scuti stars shown in Figure~7 and have 
derived reddening corrected 
$\left< V_{c,o}\right>$ mean magnitudes from the 
$\left< V_c\right>$ values adopting the 
extinction law: $A_V$=$3.1E(B-V)$. 
They are given
in Column 8 of Table~\ref{t:table2}. In the last line of the table we also 
give the corresponding values of star \#28114.
For the Kaluzny et al. stars we estimated individual reddenings from the observed
$\left< B-V \right>$ colors assuming for the stars an intrinsic color of $\left< B-V \right>_0$=0.256 mag
(see Section 3).
The best fit regression line computed 
using the de-reddened magnitudes and correcting the Kaluzny et al. variables for the 
tilt of the LMC (values in column 9 of Table~3), is: 
$$\left< V_{c,o}\right> = -2.762(\pm 0.166) \times \log P + 17.355(\pm 0.129), \,\,\,\,{\rm (8)}$$
with a rms scatter of 0.143 mag. 

In the above calculation we have adopted for OGLE II $\delta$ Scuti stars and
for star \# 28114 the reddening inferred from the RR Lyrae variables. However,  at least
in the case of star \#28114 that reddening seems too high. This is also confirmed by 
the lower reddening inferred for the Kaluzny et al. stars.
Unfortunately, we cannot derive a direct estimate of the reddening of
the OGLE $\delta$ Scuti candidates from their colors because 
 the $B-V$ color data for these stars are very poor. 
The truth is we do not have an accurate color excess for the OGLE stars, however, 
if we adopt for all the $\delta$ Scuti stars $E(B-V)$=0.08$\pm$0.02 as derived for star \#28114 and
consistently with what found for the Kaluzny et al. stars,
the $P-L$ relation becomes: 
$$\left< V_{c,o}\right> = -2.624(\pm 0.158) \times \log P + 17.526(\pm 0.123), \,\,\,\,{\rm (9)}$$ with 
a rms scatter of 0.136 mag. 
%This $P-L$ relation is the solid line in Figure~7.
Both eq.s (8) and (9) have errors in the coefficients, and dispersions
equal to or larger than those in eq. (7), thus showing that the dispersion in eq. (7)
is not due to differential reddening.

At the period of star \#28114 eq. (7) gives: $\left< V_c\right>$ = 19.974 $\pm$ 0.134 mag.
Assuming 
$E(B-V)$ = 0.08 we then obtain: $\left< V_{c,o}\right>$ = 19.726 $\pm$ 0.134 mag.
When combined with 
the $M_V$ value, $M_V$ = 1.23 $\pm$ 0.16 mag, appropriate for a $\delta$ Scuti star with 
$\log P$ = $-$0.836 and [Fe/H]=$-0.40$ (equation 1), we find: $\left< V_o\right> - M_V$ = 18.50 
$\pm$ 0.22 mag.

  We adopt as the 
``best value"  of the distance modulus to the LMC from the $P-L$ relation of the  $\delta$ Scuti stars
the mean of the solutions obtained through equation (1) with the 
apparent magnitude derived (i) from 
the value in Table~\ref{t:table1}, and (ii) by means of equation (7).
In both cases a color excess of $E(B-V)$=0.08 $\pm$0.02 was assumed.
This corresponds to: $\left< V\right>_0 - M_V$ = 18.48 $\pm$ 0.03 mag 
(dispersion of the average). This mean value was derived as the ``weighted" 
average of the individual determinations.
Weights inversely proportional to the errors of each determination
were used to calculate the mean. These errors  
take into account a 0.02  mag uncertainty in the reddening (0.06 mag in 
$A_V$).
%,
%errors in the reddening, pulsation mode definition and {\it other sources} have already 
%been taken 
%into account in the
%individual determinations. 
If a larger color excess were used, the dispersion of the average would
not change significantly.
On the other hand, systematic errors related to uncertainties 
in the pulsation mode definition and the dependence of the P-L relation on
chemical composition can be significantly large. We come back to this
point at the end of Section 5.1.
%very little if at
%all. If the dispersion is small is due to the fact that ......}. 

\subsection{A COMMON $P-L$ RELATION FOR $\delta$ SCUTI STARS AND CEPHEIDS?}

In this section we use $\delta$ Scuti stars and Cepheids in the LMC to explore the 
possibility that the two groups of variables obey the same $P-L$ relation, as first 
suggested by Fernie (1992), and recently revived by Laney, Joner \& Schwendiman (2002). 
We use both the Cepheids observed in the LMC by Laney \& Stobie (1994), as well as
the large sample of LMC Cepheids in OGLE II database (Udalski et al. 1999b).

Laney \& Stobie (1994) published $V_0$ magnitudes, periods and individual reddenings for 45
LMC Cepheids.
Their photometry was done star by star and the authors were very careful 
about eliminating systematic errors. The Laney \& Stobie's $V_0$ magnitudes were reddened back to the 
original observed numbers utilizing the author's color excesses and assuming $A_{V}$=$3.1E(B-V)$.
Magnitudes were also corrected to account for the tilt of the LMC applying corrections given 
in the Laney \& Stobie paper. 
The best fit regression line computed on these 45 Cepheids is:
$$\left< V \right> = -2.764(\pm 0.114) \times \log P + 17.380(\pm 0.151), \,\,\,\,{\rm (10)}$$ with 
a rms scatter of 0.272 mag. 

According to Udalski et al. (1999b) the $P-L$ relation of the LMC Cepheids in the OGLE II 
catalogue is given by the equation: 
$$\left< V_0\right> = -2.760(\pm 0.031) \times \log P + 17.042(\pm 0.042), \,\,\,\,{\rm (11)}$$
with a rms scatter $\sigma$=0.159.
This relation was obtained from a sample of 649 fundamental mode LMC Cepheids
with periods in the range from $\sim 2.5$ to $\sim 32$ d, correcting for 
an average reddening $E(B-V)$ = 0.147 mag, and assuming 
the extinction law: $A_V$=$3.24E(B-V)$.  
%We notice that 
The slopes of both these two relations are   
in excellent agreement with the theoretical slope of $-2.75\pm0.02$ predicted by nonlinear 
convective models of Cepheids with $Z=0.008$, 
$Y=0.25$ and $\log P \le 1.5$
(see Caputo, Marconi, \& Musella 2000, 2002).

The $P-L$ relations of $\delta$ Scuti stars and Cepheids in the LMC are compared 
in Figure~8,
where we plot the 
$\left< V_c\right>$ magnitudes versus $\log P$ of the 32 LMC fundamental/fundamentalized $\delta$ Scuti 
candidates analyzed in Section~3.1 and their $P-L$ relation (equation (7), solid line), along with 
an extrapolation of the $P-L$ relations defined by the
Laney \& Stobie's Cepheids (equations (10), long-dashed line)
and Udalski et al.'s Cepheids (equation (11), short-dashed line). 
The extrapolations are over approximately 5 mag. The Cepheid relations extend over the $\log{P}$ interval $0.4 < \log{P} < 1.9$ and the $\delta$ Scuti
stars over the range $-1.2 < \log{P} < -0.55$.
A correction of 
$A_V$=$3.24E(B-V)$=0.476 mag has been added to the $\left< V_0\right>$ values calculated from eq. 
(11) to obtain $\left< V\right>$ values, that were then transformed to DF05 photometric zero-point.
Since we have no stars in common to check systematics in magnitude scales we assumed that 
 the Laney \& Stobie's magnitudes are on the $V_c$ system.
%The three $P-L$ relations run parallel.
The difference in magnitude at $\log P$ = $-0.80$ given 
by the three equations is: 0.29 mag with Laney \& Stobie's Cepheids, and only 0.09 mag with OGLE II's
Cepheids.  The latter result gives some support to
the original 
suggestion of Fernie (1992) that the two groups of variables may fit the same $P-L$ relation.  

To investigate this point further we have computed a common $P-L$ relation 
utilizing the sample of 32  $\delta$ Scuti stars and the 45 Cepheids by Laney \& Stobie (1994).
The corresponding best fit regression line is:
$$\left< V \right> = -2.894(\pm 0.025) \times \log P + 17.556(\pm 0.029), \,\,\,\,{\rm (12)}$$ with 
rms scatter of 0.230 mag. Notice that the slope is similar to the
slope in equation (1), and the very small errors in the slope and intercept.
This $P-L$ relation is the solid line in Figure~9.
The open squares are points calculated from the Udalski et al.'s $P-L$ relation 
(equation (11)) respectively at $\log P$=0.45; 0.55; 0.65;  0.95 and 1.45, 
reddened back by
adding $A_V$=0.476 mag, and correced to DF05
photometric zero point. 
They fit well  the common $P-L$ relation described by eq. (12).
Equation (12) has a sizeable rms scatter of 0.23 mag. This is caused by the inclusion of the 
Cepheids in the solution. The cosmic scatter in the Cepheid $P-L$ relation is approximately 0.24 mag which
compares with the cosmic scatter in the $\delta$ Scuti relation of approximately 0.08-0.10 mag. The cosmic
scatter is due to the finite width of the instrability strip which is smaller for the $\delta$ Scuti stars
than for the Cepheids (compare the $\pm$0.134 mag of Eq. (7) to the 0.23 mag rms of the $\delta$ Scuti stars 
plus Cepheids solution in Eq. (12)).

We lack objects in the region around $\log P \sim $0.00. This is where Anomalous
Cepheids (ACs) are generally found. Only a few AC candidates are known in the LMC (DF05), 
they are shown by asterisks in Figure~9.  

Note that if the original suggestion of Fernie (1992) is correct, that the Cepheids and 
$\delta$ Scuti stars obey a similar, unique $P-L$ relation, it implies that the $P-L$
relation is valid over a range of 10 magnitudes (ratio of brightness 10,000 to 1), 
since $\delta$ Scuti stars become as faint as $M_V$=3.0 mag and Cepheids can reach 
$M_V=-7.0$ mag.  Our analysis indicates that if Fernie's suggestion is not correct, 
it must be very nearly so.  When both Cepheids and $\delta$ Scuti stars are available 
it should be possible to improve the slope and possibly also the intercept 
of the $P-L$ relation over what one can obtain by using just one set of variables.
 
\section{THE $\delta$ SCUTI ABSOLUTE MAGNITUDE FROM THE LIGHT CURVES 
MODEL FITTING}

A powerful  method to derive the distance and the intrinsic
stellar parameters of observed pulsating stars using the output of
nonlinear convective hydrodynamic models is the direct comparison
between the theoretical and the empirical light (and radial velocity)
curves. This procedure has been successfully applied to both  RR Lyrae
(Bono, Castellani, \& Marconi 2000, Di Fabrizio et al. 2002,
Castellani, Degl'Innocenti, \& Marconi 2002, Marconi \& Clementini 2005)
and Classical Cepheids (Wood, Arnold \& Sebo 1997, Bono, Castellani \& 
Marconi 2002, Keller \& Wood 2002). In particular, by modeling the $BV$
light curves of 14 LMC RR Lyrae from the database by DF05,  
Marconi \& Clementini (2005) find an average distance
modulus for the LMC of  $\mu_{0}$= 18.54 $\pm$ 0.02 ($\sigma$= 0.09
mag, standard deviation of the average), consistent with the results
obtained with the same method applied to LMC Classical Cepheids (Bono,
Castellani, Marconi 2002, Keller \& Wood 2002).  In this section we
present the theoretical fit of the light curve of the $\delta$
Scuti star discovered in the LMC by C03 and DF05. This is the first
application of the method to a member of the  $\delta$ Scuti class, providing 
a further constraint on the distance  to the LMC, as well as an
additional test of the predictive capability of the current theoretical
scenario.  

At variance with more evolved pulsating stars, $\delta$
Scuti models are very time consuming, due to the low pulsation growth
rate, typical of relatively massive, low luminosity
structures. Therefore, the construction of an extended and detailed atlas
of theoretical light curves is quite hard. Bono et al. (1997) on the
basis of a sequence of nonlinear convective models at fixed mass and
luminosity found that the first three radial modes are excited in these
pulsators and that, at variance with current empirical
classifications,  moving from the second overtone  to lower
pulsational modes the amplitudes  progressively decrease and the shape
of the light curves becomes more sinusoidal, with low-amplitude
pulsators, like our LMC $\delta$ Scuti, either pulsating in the
fundamental or in the first overtone mode.  On this basis we tried to
model the observed light curves shown in Fig. 1 both with fundamental
and first overtone models with  chemical composition $Z=0.008$ 
(corresponding to [Fe/H]=$-$0.40), $Y=0.25$. These 
abundances were chosen  
on the basis of current empirical estimates for the intermediate age stars
in the LMC (Russell \& Dopita 1990, Luck et al. 1998). Once the 
chemical composition is fixed, 
we adopted a stellar mass in the range 1.5-2.5
$M_{\odot}$, while  the luminosity and the effective temperature were varied in
order to reproduce the observed period. Among these isoperiodic model
sequences we selected the model which was able to best reproduce the
amplitude and the morphology of the 
observed light curves. 
 We do not
find any fundamental mode model satisfying the above requirements, whereas
our first overtone best fit model is shown in Figure~10
and
corresponds to a post-MS $\delta$ Scuti model with $M=2.0M_{\odot}$, $\log{L/L_{\odot}}=1.41$ 
and $T_e=$6900 $K$. The corresponding intensity-weighted absolute mean magnitude is 
$\left< M_V \right>$=1.21$\pm$0.10, where the error includes the uncertainty of the
model fitting (0.05 mag), which was evaluated by varying the stellar parameters around
the best-fit values, and the photometric uncertainty (0.09 mag). 
We remind that this value is obtained directly from the model fitting, 
without requiring the knowledge of the distance modulus. 
We note that the model light curve in Figure~10 shows a bump before the maximum light,
that is not clearly seen in the observed curves. However, observations are too scattered
to definitely confirm or rule out its actual existence.
Combining the $\left< M_V \right>$ value from the model fitting with 
the star apparent magnitude $\left< V \right>$=19.94 $\pm$ 0.09 mag, and
adopting $E(B-V)$=0.08$\pm$0.02 and $A_V=3.1E(B-V)$, consistently with what suggested by the star
intrinsic color, we obtain an absolute distance modulus for the LMC of
18.48$\pm$0.15 mag, that agrees very well with the results from the $P-L$
relations of Galactic and LMC $\delta$ Scuti stars.

By matching the light curves in $B$ and $V$ we 
infer the two apparent  distance moduli $\mu_B$=18.77$\pm$0.10 and $\mu_V$=18.70$\pm$0.10,
where errors include both the photometric uncertainties (by far the largest contributions) 
and the errors in the model 
fitting procedure. Within their error bars the apparent distance moduli are fully consistent with the
$\left< M_V \right>$ value derived above.
 The difference between these two 
quantities allows us to derive a ``pulsational"
reddening of 0.07$\pm$ 0.14 mag.
 Although affected by large uncertainty, this reddening is 
consistent with the
low reddening inferred from the star intrinsic color, and 
is about 0.05 mag smaller than the average 
reddening derived by C03 from the RR Lyrae stars in the area.
The $I$ light curve of the star has fewer data points than the $V$ and $B$ ones, 
nevertheless, we find that 
our model best-fitting the $V$ and $B$ light curves also reproduces the 
%our best-fit model also reproduces 
amplitude of the $I$ band light curve, and leads to an apparent distance modulus in $I$  which is consistent 
with the above quoted ``pulsational" color excess.
%and that the inferred  distance modulus is consistent with the assumed color excess.

\section{SUMMARY AND DISCUSSION}

The absolute magnitude of the LMC $\delta$ Scuti star \#28114 in DF05 catalog
was determined from the $P-L$ relation of the Galactic $\delta$ Scuti stars (equation 1) 
and from the
theoretical modeling of the star light curves (Section 4), deriving values of
$M_V$=1.23$\pm 0.15$ and 1.21$\pm 0.10$ mag, respectively.
Combined with the star apparent magnitude $V$=19.94 $\pm$ 0.09 mag and assuming for the 
reddening
$E(B-V)$=0.08 $\pm 0.02$ (which is the average of the ``pulsational" reddening:
$E(B-V)$=0.070, the color excess inferred from the star intrinsic
color: $E(B-V)$=0.076, and the LMC foreground reddening: $E(B-V)$=0.075, according to
Schlegel et al. 1998 maps), the corresponding distance moduli for the
LMC are: 18.46$\pm$0.19 and 18.48$\pm$0.15 mag respectively.
The star was used along with 24 $\delta$ Scuti candidates discovered in the 
OGLE II survey of the LMC, and 7 $\delta$ Scuti stars discovered in the LMC by
Kaluzny et al. (2003, 2006), to define the $P-L$ relation of the LMC $\delta$ Scuti stars:  
$\left< V_{c}\right> = -2.635 (\pm 0.156) \times \log P + 17.770 (\pm 0.121), \sigma=0.134$, which is tied
to DF05 photometric zero-point. For $E(B-V)$=0.08$\pm$0.02 mag this relation
leads to $\left< V_0\right> - M_V$ = 18.50 $\pm$ 0.22 at the period and metallicity
of \#28114. The final distance modulus of the LMC derived from the $\delta$ Scuti stars
is then: 18.48$\pm$0.02, where the error is the standard deviations of the weighted average of the 
three above solutions.
    
\subsection{COMPARISON WITH OTHER PULSATING VARIABLES}
   
Thanks to its composite stellar population the LMC hosts different types of pulsating stars 
which are standard candles and that, once distortions and depth effects due to the LMC geometry
are properly taken into account, can 
%all being at the same distance from us, 
provide the opportunity of a direct and easy cross check of the predicted distances.
In the following we will compare our results from the LMC $\delta$ Scuti variables with
the distance inferred from the other two major types of pulsating variables found in the LMC, 
namely: the RR Lyrae stars, and the classical Cepheids.    

{\it -RR Lyrae stars} 

C03 found that the average $\left< V_0\right>$ magnitude of the RR Lyrae 
stars in two regions of the LMC containing a total number of 108 of these variables, for stars with 
[Fe/H] = $-$1.5 is: $\left< V_0\right>$ = 19.05$\pm$0.06 mag. 
Individual reddenings of the RR Lyrae stars in each region were found that led to this accurate 
$\left< V_0\right>$ magnitude. 
 Cacciari \& Clementini (2003) suggest the best $M_V(RR)$ value at 
[Fe/H] = $-$1.5 is, $M_V(RR)$ = 0.59 $\pm$ 0.03.
 Thus the distance modulus is $\left< V_0\right> - M_V$ = 18.46 $\pm$ 0.07 mag.  
 
Soszy\'nski et al. (2003) give $\left< V(RRab)\right>$ = 19.36$\pm$0.03 mag 
and $\left< V(RRc)\right>$ = 19.31$\pm$0.02 mag from the OGLE~II sample of RR Lyrae stars in
the LMC.
These values transformed to DF05 photometric zero point give
19.42 and 19.37 for {\it ab-} and {\it c-} type RR Lyrae respectively, and an average of
$\left< V(RR)\right>$=19.40$\pm$0.04 (standard deviation of the average).
Adopting C03 reddening scale for the LMC RR Lyrae stars that, on average, is 0.03 mag
lower than OGLE's, and Cacciari \& Clementini (2003) $M_V(RR)$, again we find 
$\left< V_0(RR)\right>$=19.05 and $\left< V_0\right> - M_V$ = 18.46 $\pm$0.07 mag 
for the distance modulus of the LMC, in perfect agreement with  C03 solution.

Marconi \& Clementini (2005) derive $\left< V_0\right> - M_V$ = 18.54 $\pm$0.09 mag from the
theoretical modeling of the light curves of LMC RR Lyrae stars. We will thus adopt for the
distance modulus from the RR Lyrae stars the weighted average of the above solutions, 
namely: 18.49 $\pm$0.06 (standard deviation of the weighted average).

{\it Cepheids}

Next we turn to the Cepheid variables.  We will make our computation at $\log P$ = 0.55 which 
is approximately at the centroid of where the largest number of these variables are found, 
and will use Udalski et al. (1999b) $P-L$ relation for Cepheids (equation (11)), since it is based 
on a very large number of stars and is the relation most frequently used in the recent literature.  The 
$\left< V\right>_0$ magnitude at $\log P$ = 0.55, given by equation (11) is: 
$\left< V\right>_0$ = 15.524, that corrected back to an original $\left< V\right>$ 
magnitude by adding  $A_{V}$ = 0.476 mag and transformed to DF05 
photometric zero point, becomes: $\left< V \right>$ = 16.035 mag.  
At $\log P$ = $-$0.75 and [Fe/H]=$-0.40$  the $\delta$ Scuti stars give 
$\left< V\right>$ = 19.764 $\pm$ 0.131 mag (equation 7) and 
 $M_V$ = 0.981 $\pm$ 0.16 mag (equation 1).
Combining differences, on the assumption of the same apparent distance modulus 
($\mu_V$=18.783 $\pm$0.207) for the LMC $\delta$ Scuti stars and Cepheids, yields 
$M_V$ = $-2.748 \pm 0.262$ mag for the 
Cepheids.
Again we assume the color excess is 0.08$\pm$ 0.02 mag and the absorption in $V$ is 
$A_{V}$ = 0.248 mag.  
We find a distance modulus of $\mu_0$=$\mu_V$-$A_{V}$= 18.535 $\pm 0.216$ mag.  In this solution we 
have utilized the 
$\delta$ Scuti stars to infer the luminosities of the Cepheids.
The above distance modulus is in excellent agreement with 
that obtained by an independent Cepheid-based method, namely the theoretical fitting of the observed 
light curves of the LMC Cepheids: 18.55$\pm$0.09 (Keller \& Wood  2002, 2006); 
18.52$\pm$0.06 (Bono, 
Castellani, Marconi 2002). We adopt for the Cepheids the weighted average of the above three
solutions, namely: 18.53 $\pm$0.02 (standard deviation of the weighted average).

The very good agreement found among results from the model-fitting 
of the observed
light curves of the LMC Cepheids, RR Lyrae and
$\delta$ Scuti stars (see Sect. 4 of the present paper), demonstrates the reliability
of this method, and its marginal dependence on the adopted pulsation code, as well as on 
physical and numerical assumptions in the model calculations.

In summary, our best values for the distance modulus of the
LMC from the three different types of pulsating variables are:
18.48 $\pm$ 0.02 mag from the $\delta$ Scuti stars; 
18.49 $\pm$ 0.06 mag from the RR Lyrae stars;
18.53 $\pm$ 0.02 mag from the Cepheids.
Here the errors are the standard deviations of the averages 
of independent solutions. Systematic errors related to reddening uncertainties and, for 
the delta Scuti stars considered in this paper, uncertanties in the 
pulsation mode definition, dependence of the $P-L$ relation on chemical composition, pulsation 
and/or model atmosphere  
limitations, are difficult to quantify and can be significantly large (up to a couple of tenths 
of magnitude). 
%Improvement upon both reddening and pulsation mode issues  are expected from 
New data for the LMC delta Scuti stars from the
SuperMACHO survey of the LMC (Cook 2006, private communication) enabling a direct definition 
of 
periods and pulsation modes, and reddening determinations from the star's intrinsic 
colors should in the future allow to cut down significantly these error contributions, 
thus streghtening the delta Scuti star solution.
However, and notwhitstanding the uncertainties still present, our finding that the three 
groups of 
pulsating stars give 
consistent distance moduli for the LMC already reprents a remarkable result,
probably traceable to the use of the proper color excess and to the application of 
independent techinques
to the various types of variables. 
In particular, all three groups of pulsating stars
give distance moduli that are consistent with the ``long" astronomical distance scale for the
LMC, in very good agreement with the value of 19.52 $\pm$ 0.10 mag derived by C03 using a 
variety of different distance
indicators of both Population I and II and a careful treatment of the error sources.
%%Still, we find that our preliminary results, 
%%are ...confortably ... and agree very well with ....
%%nel C03 che danno come media da tanti metodi diversi 19.52 $\pm$ 0.10 mag.
%%
%%is a remarcable result 
%%
%%, where the errors are the standard deviations of the averages 
%%of independent solutions.
%%{\it Systematic errors, related to reddening uncertainties, dependence on chemical composition, pulsation and/or atmosphere model limitations, ecc... can be significantly larger.} 
%%
%%{\bf ( Qui il referee chiede di: adding a few sentences describing the error sources and what they mean in
%%terms of the `actual' error in the 18.49 number.
%%Forse l'unica discussion che si puo' ancora fare qui e' quella del reddening e della definizione del modo di
%%pulsazione, la necessita' di avere piu' dati, piu' accurati in modo da poter anche ricavare il reddening
%%direttamente dai colori intrinseci delle $\delta$ Scuti e poi forse si puo' mettere il confronto con le stime
%%nel C03 che danno come media da tanti metodi diversi 19.52 $\pm$ 0.10 mag.}
%%
%% 
%%The three groups of pulsating stars give very similar distance moduli, 
%%this can probably be traced to the use of the proper color excess for the 
%%various types of variables, and to multiple solutions. Moreover, all three groups of 
%%pulsating stars
%%give distance moduli that are consistent with the ``long" astronomical distance scale for the
%%LMC.
%%
%%{\bf and are consistent with what found by C03 19.52 $\pm$ 0.10 mag using a variety of different distance
%%indicators of both Pop. I and II.}

In conclusion, let us emphasize the desirability of identifying other 
$\delta $ Scuti stars in the Magellanic Clouds. In the SMC they should be found 
between $\left< V \right> \approx 21.6$ mag at log $P \approx -1.30$ to 
$\left< V \right> \approx 19.9$ mag at log $P\approx -0.86$.
These variables can be also identified in other Local Group galaxies,
provided that deep enough observations are obtained.  For instance,  
 a large number of SX Phe stars have been successfully identified 
in the Fornax dwarf spheroidal galaxy at 22  $< \left< V \right> <$ 24 mag  
(Clementini et al. 2004, Poretti et al. 2006a,b). 
Not only 
do the variables in these galaxies allow
 to improve our detailed knowledge of the 
properties of the variables, but distances to the galaxies as well. Note 
that the LMC, the SMC, Fornax, as well as other galaxies close-by provide 
the opportunity to directly and easily compare 
luminosities of the three 
pulsating variables: $\delta $ Scuti, RR Lyrae and Cepheids, that occupy the 
instability strip.  The present paper is a first attempt in this direction.

\acknowledgments

We thank Mario Mateo and Ennio Poretti for informations on the dwarf Cepheid
statistics outside the Milky Way and an anonimous referee for his/her comments and
suggestions. 
Financial support for this study was provided by MIUR, under the scientific
project 2004020323, (P.I.: Massimo Capaccioli).

\clearpage
 
\begin{figure*}
\includegraphics[width=18cm, bb=18 519 593 719]{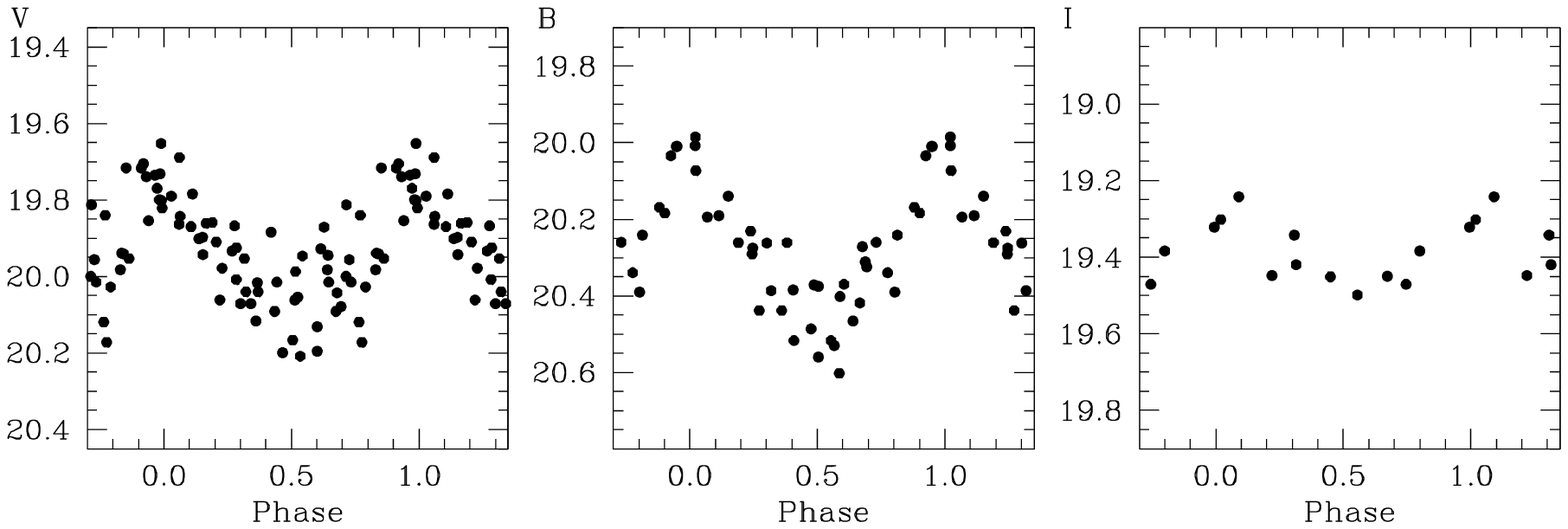} 
\figcaption{$V,B,I$ light curves of the program star: DF05\_LMC\_A \#28114.}
\label{f:fig1}
\end{figure*} 

\clearpage
 
\begin{figure*}
\includegraphics[width=18cm, bb=18 144 592 719]{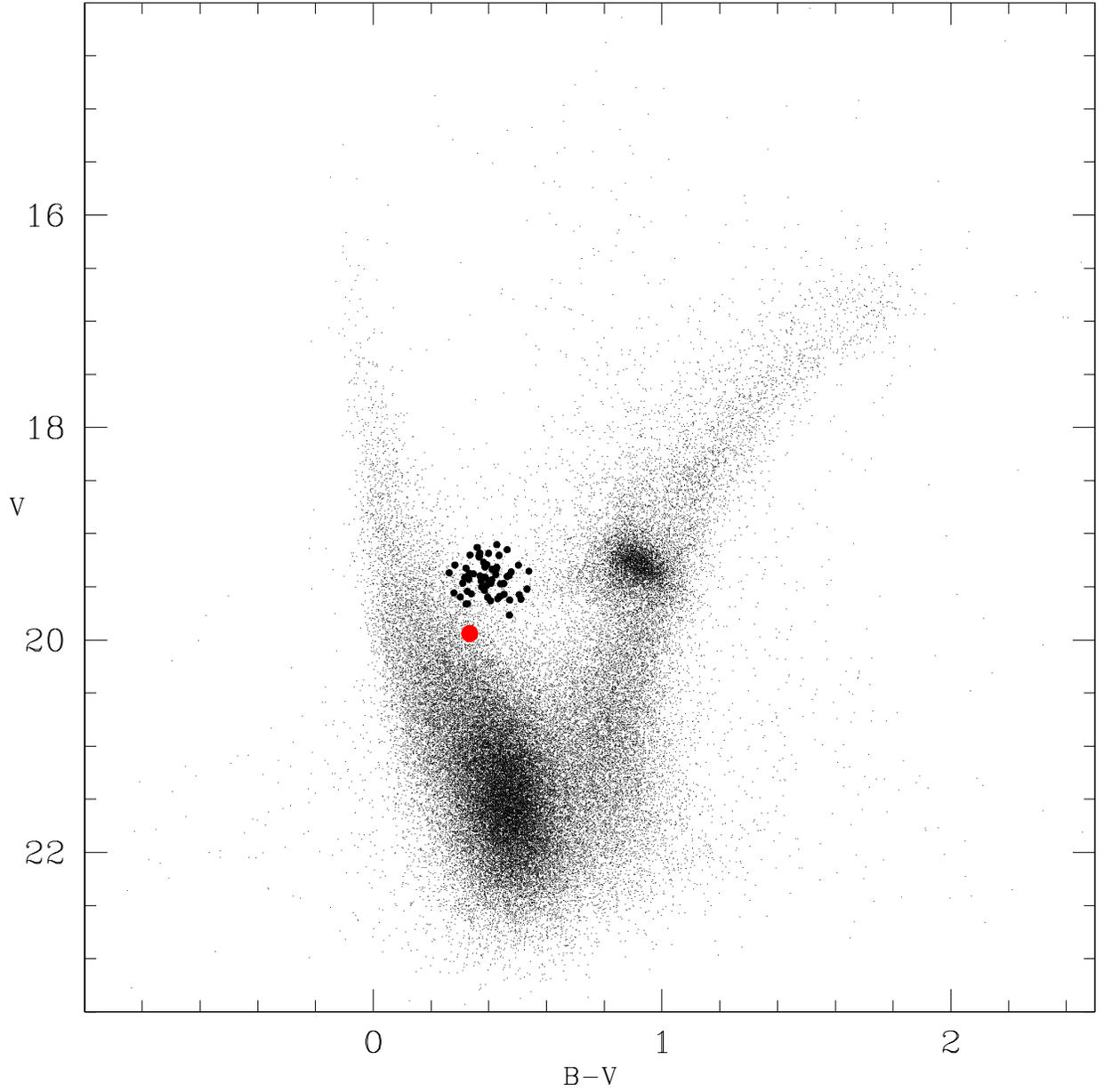} 
\figcaption{Position of the $\delta$ Scuti star (big filled circle) in the color magnitude diagram 
of the LMC Field A of DF05. The small filled circles are RR Lyrae stars. The variables are 
plotted according to their intensity-averaged magnitudes and colors.}
\label{f:fig2}
\end{figure*} 
\clearpage
 
\begin{figure*}
\includegraphics[width=18cm, bb=18 144 592 719]{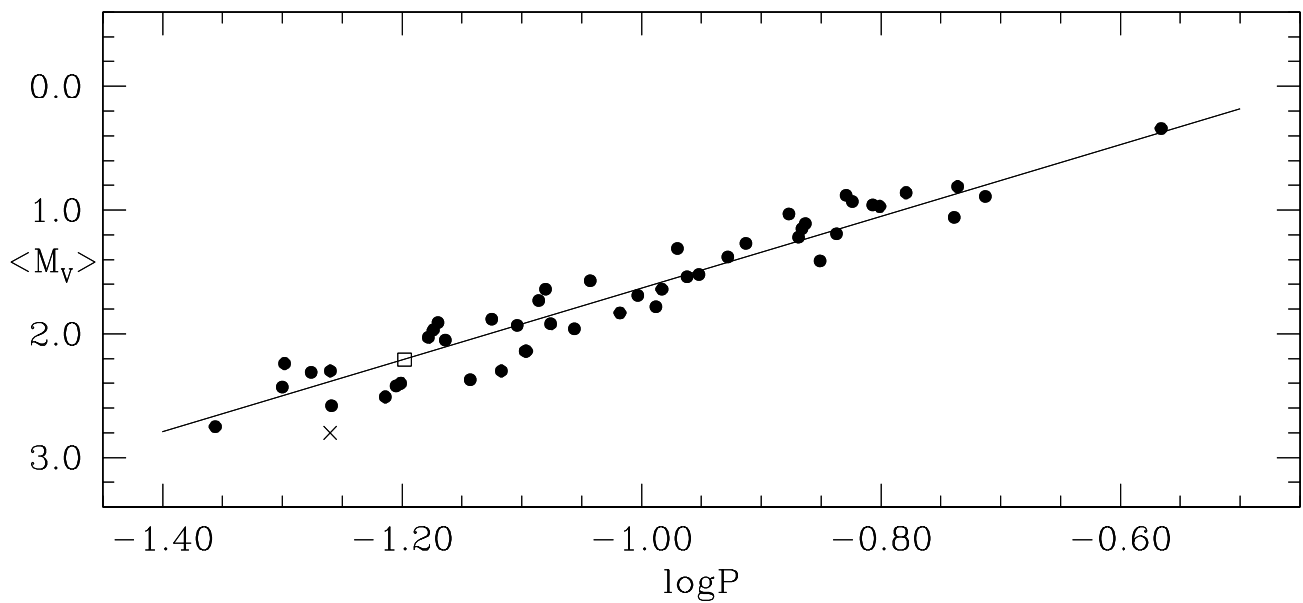}
\figcaption{Period-luminosity relation of Galactic $\delta$ Scuti stars. The $\delta$ Scuti stars shown are
metal-strong variables ($m1 > 0.155$) with parallax errors $\leq$ 0.10 mas. The line is equation (1): 
$M_V=-2.90 \times \log P-1.27$ (the metallicity term, $-0.19 \times $[Fe/H], set equal to zero). The open square is the star
Altair, and the cross is the metal-poor variable ([Fe/H]$\sim -1.5$) SX Phe. It is fainter than the 
metal-strong stars of similar period. 
%One half the length of the vertical line in the cross (0.13 mag) is
Typical errors in the $M_V$ values of the stars plotted are $\pm 0.13$ mag. They are due to 
parallax 
errors.  
}
\label{f:fig2}
\end{figure*} 

\clearpage
 
\begin{figure*}
\includegraphics[width=18cm, bb=18 144 592 719]{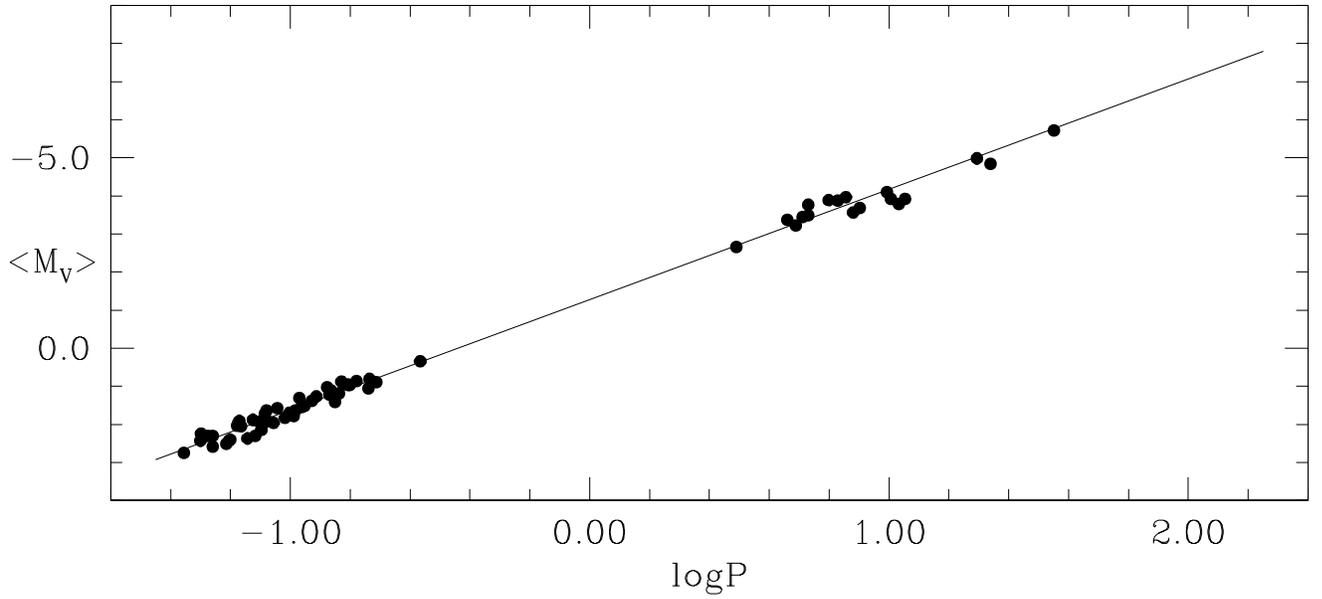} 
\figcaption{Fit of equation (1) to Galactic $\delta$ Scuti and Cepheid variable stars.
The $\delta$ Scuti stars are the same as shown in Fig.~3. The Cepheid's $M_V$ values are 
from Kervella et al. (2004) and Hoyle et al. (2003). The line is equation (1) 
%$M_V=-2.90 \times \log P-1.27$ 
with the metallicity term set
equal to zero.
}
\label{f:fig2}
\end{figure*}

\clearpage
 
\begin{figure*}
\includegraphics[width=16cm, bb=18 144 592 430]{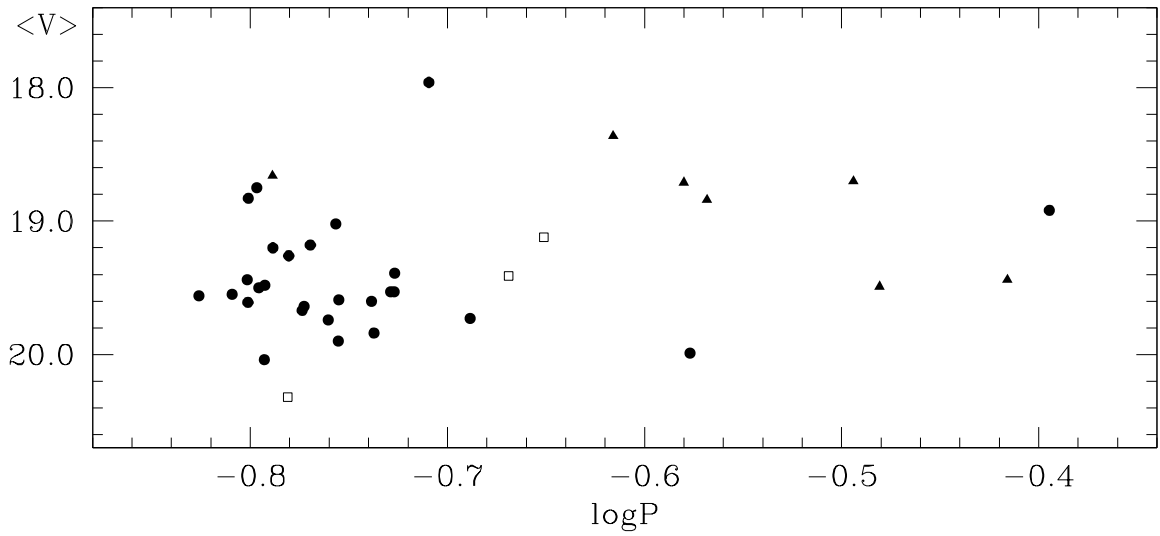} 
\figcaption{$\left< V\right>$ versus $\log P$ distribution 
of the 37 OGLE II short-period stars, classified as ``other 
pulsating variables" in Soszy\'nski et al. (2003).
The $\left< V\right>$ magnitude 
is the observed intensity-averaged magnitude. Filled circles
are single-mode pulsators; filled triangles are double-mode pulsators; open squares are
stars vibrating in two periods in which the period ratio is $P_{1}/P \sim $0.76.
For all the double-mode pulsators the shortest periods are used in the plot.
}
\label{f:fig3}
\end{figure*} 

\clearpage
 
\begin{figure*}
\includegraphics[width=16cm, bb=18 144 592 430]{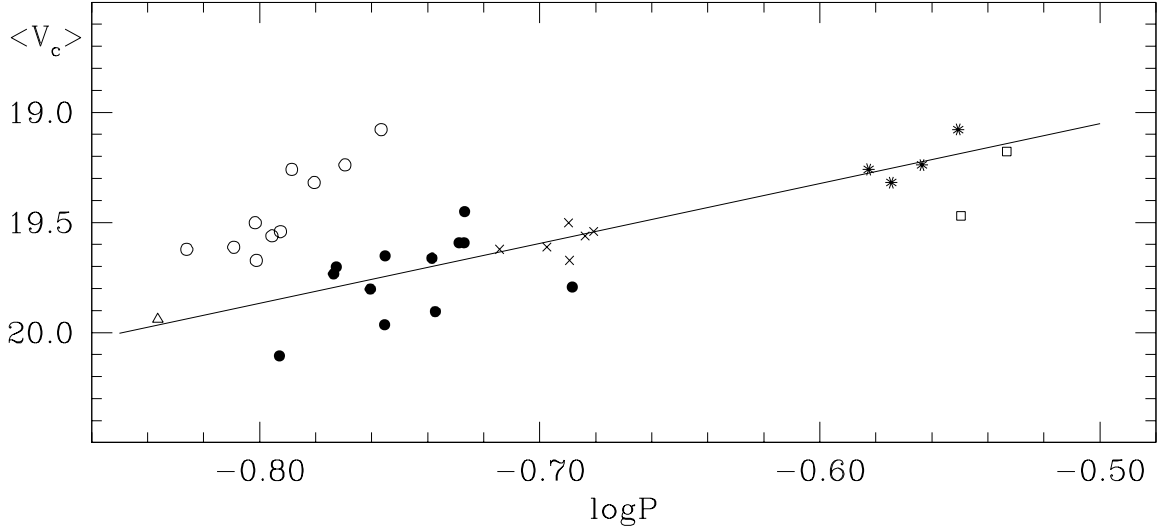} 
\figcaption{$V$- band $P-L$ relation of the OGLE II LMC $\delta$ Scuti candidates 
using magnitudes corrected to DF05 photometric zero-point (see text).  
Solid dots are fundamental mode pulsators. The squares 
represent stars vibrating in two periods in which the period ratio is $P1/P \simeq$ 0.76,
they are plotted according to their longest periods (see Column 4 of Table~\ref{t:table2}).   
The open triangle represents the position of the star \#28114 discussed in the text, according to
the fundamentalized period.  
The open circles 
are stars pulsating in the first harmonic or higher modes. Among them the
group of six stars
 with 19.7$<\left< V\right><$ 19.4 mag and $-0.84<\log P < -0.79$ which must be first overtone pulsators 
were fundamentalized assuming $P_{1}/P = $0.773 and also plotted as X's. 
 The four stars at 19.0$ <\left< V_c\right> <$ 19.4 mag and $\log P \sim -0.77$, are likely 
pulsating in the second overtone, they were fundamentalized assuming $P_{2}/P = $0.6223 and 
are also plotted as asterisks. The solid line is the best fit regression line directly 
computed on the fundamental/fundamentalized stars, its equation
is $\left< V\right> = -2.577(\pm 0.314) \times \log P + 17.814 (\pm 0.219), \sigma=0.125$ (eq. (5) in 
the text).
}
\label{f:fig4}
\end{figure*}

\clearpage

\begin{figure*}
\includegraphics[width=16cm, bb=18 144 592 430]{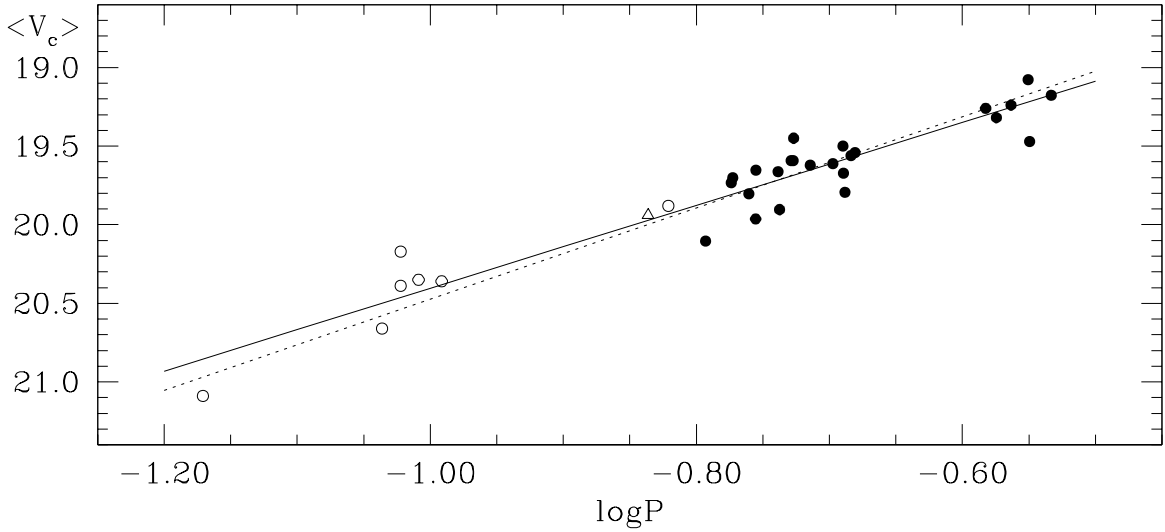} 
\figcaption{$V$- band $P-L$ relation of the 32 LMC $\delta$ Scuti candidates. 
Fundamental/fundamentalized periods are used for all of them.
Filled circles are OGLE II $\delta$ Scuti candidates (24 stars), open circles
are Kaluzny et al. (2003, 2006) stars (7 objects). We have assumed that the 
Kaluzny et al. magnitudes are on the Vc system and have corrected their magnitudes to 
account for the LMC tilt. The open triangle
is star \#28114. The dashed line is
the $P-L$ relation of the Galactic $\delta$ Scuti stars with  
[Fe/H]$\geq -1.5$ (equation (1)) with the metallicity term set equal to zero 
%computed at [Fe/H]=$-0.40$ 
and shifted to 
fit the LMC $\delta$ Scuti stars, its equation is:
$\left< V\right> = -2.90 \times \log P + 17.574$.
The best fit regression line directly computed on the stars is shown by the
solid line, its equation is: $\left< V\right> = -2.635 (\pm 0.156) \times \log P + 
17.770 (\pm 0.121)$,
$\sigma$=0.134 mag (eq. (7) in the text).}
\label{f:fig5}
\end{figure*}

\clearpage

\begin{figure*}
\includegraphics[width=16cm, bb=18 144 592 430]{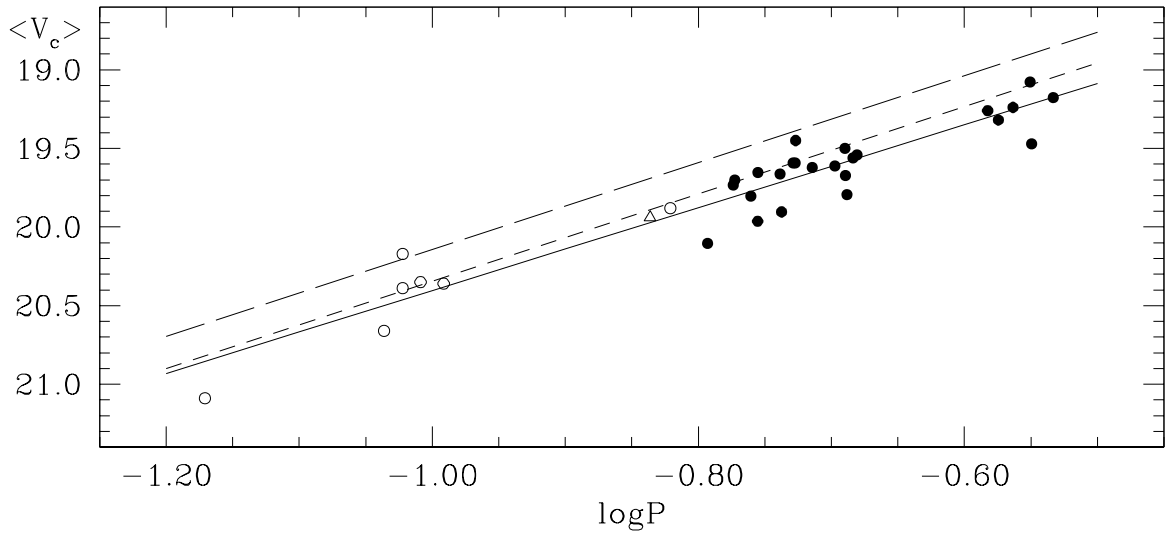} 
\figcaption{Comparison between the LMC $\delta$ Scuti $P-L$ relation 
(equation (7), solid line) and the $P-L$ relations defined by Cepheids in Laney \& Stobie's 
(equation (10), long-dashed line) and Udalski et al.'s (from equation (11) reddened back
and transformed to DF05 photometric zero point, short-dashed line) samples, 
respectively.}
\label{f:fig6}
\end{figure*}

\clearpage

\begin{figure*}
\includegraphics[width=16cm, bb=18 144 592 430]{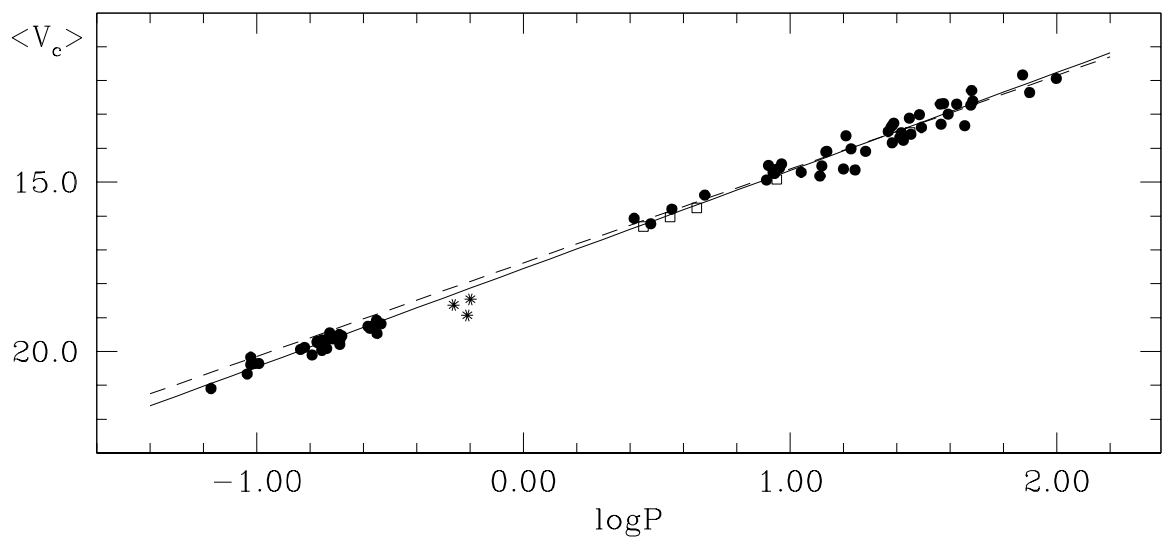} 
\figcaption{Solid line: common $P-L$ relation computed from the LMC $\delta$ Scuti stars
and Laney \& Stobie's  Cepheids, its equation is: 
$\left< V_c\right> = -2.894 (\pm 0.025) \times \log P + 
17.556 (\pm 0.029)$,
$\sigma$=0.230 mag (eq. (12) in the text).
Open squares are points computed from OGLE II $P-L$ relation (equation (11)), 
reddened back
and transformed to DF05 photometric zero point. Asterisks are candidate
Anomalous Cepheids in the LMC from DF05. The dashed line is the Laney \& Stobie's Cepheid
$P-L$ relation (equation (10)).}
\label{f:fig7}
\end{figure*}

\clearpage
 
\begin{figure*}
\includegraphics[width=18cm, bb=18 144 592 719]{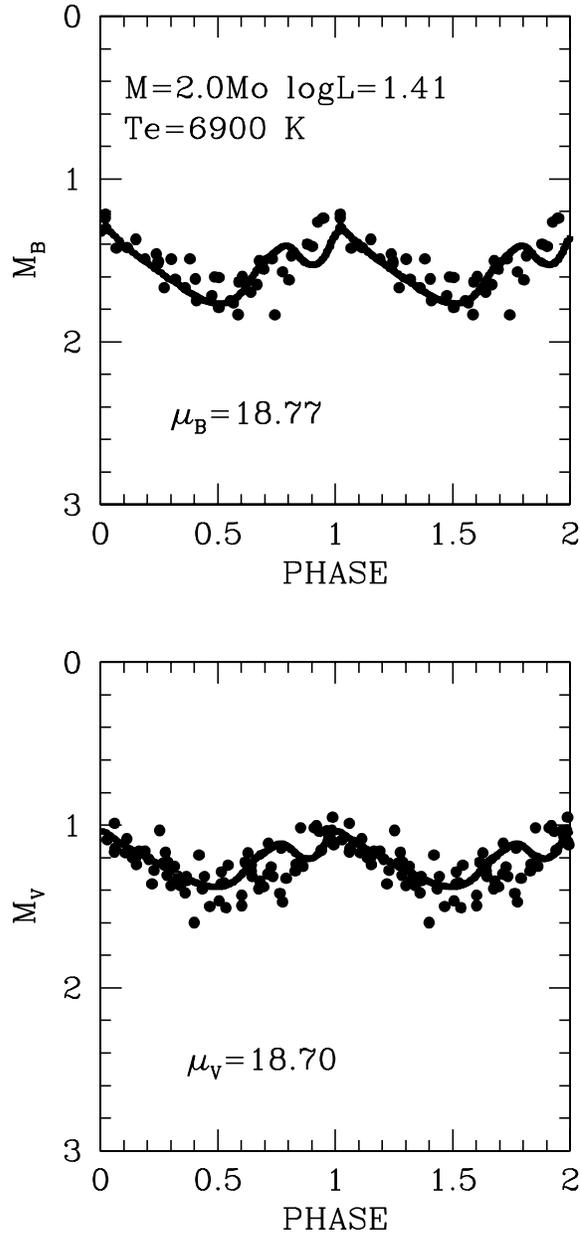} 
\figcaption{Results from the theoretical modeling of the $B$ (upper) 
and $V$ (lower) light curves of the LMC $\delta$ Scuti star \#28114.}
\label{f:fig8}
\end{figure*} 

\clearpage

\begin{table*}[ht]
\caption{Observed characteristics of the $\delta$ Scuti star}
\begin{center}
\small
\begin{tabular}{rccccccccccc}
\hline
\hline
Name~~&Epoch &$P_1$    & $\left< V \right>$&$\left< B \right>$&$\left< I \right>$&
$\left< B \right> - \left< V \right>$&$\left< V \right> - \left< I \right>$&$A_{V}$ &$A_{B}$&
$A_{I}$ \\
    &	$-$2451183&  
                    &            &           &           &        &       &\\
\hline
28114&0.63220&0.112677&19.939& 
20.271& 19.393&0.332&0.546&0.283&0.399&0.195 \\
\hline
\end{tabular}
\label{t:table1}
\end{center}
\end{table*}
\normalsize
Notes:\\
Star identifier, first harmonic period and epoch (corresponding to the time 
of maximum light) are from DF05.\\
$\left< B \right>$,$\left< V \right>$,$\left< I \right>$ magnitudes 
are intensity-averaged mean values, from DF05.\\  

\clearpage

\begin{table*}[ht]
\caption{$\delta$ Scuti candidate stars in the OGLE II survey of the LMC along with DF05 LMC $\delta$ Scuti star.}
\begin{center}
\tiny
\begin{tabular}{lcrlccccc}
\hline
\hline
LMC field & Star ID& Star number & ~~~~~~~~~~~$P$&$\left< V \right>$&$\left< V_o \right>$& $\left< V_c \right>$&$\left< V_{c,o} \right>$&Remarks\\
~~~~~S03   & S03& S03~~~~~~~     & ~~~~~~~~~S03          &  S03& S03  &DF05 & C03   &               \\
          &        &             &                  &         & redd.&phot. z.point& redd.& \\  
\hline
LMC\_SC2 & OGLE053130.51-701842.6& 220195 & 0.1826207&	   19.60& 19.20&  19.662& 19.368&  (1) \\   
LMC\_SC3 & OGLE052944.24-695341.7& 388188 & 0.1830890&     19.84& 19.45&  19.904& 19.619&  (1) \\  
LMC\_SC4 & OGLE052604.28-700419.7& 148159 & 0.1757183&     19.59& 19.17&  19.652& 19.337&  (1) \\ 
LMC\_SC5 & OGLE052437.21-692549.1& 442106 & 0.1735753&     19.74& 19.31&  19.803& 19.479&  (1) \\ 
LMC\_SC7 & OGLE051750.12-692500.3&  60291 & 0.1610768&     20.04& 19.59&  20.105& 19.762&  (1) \\ 
LMC\_SC7 & OGLE051911.43-693946.1& 266085 & 0.1867372&     19.53& 19.07&  19.592& 19.238&  (1) \\ 
LMC\_SC7 & OGLE051937.08-693746.7& 391110 & 0.1875760&     19.39& 18.94&  19.450& 19.107&  (1) \\ 
LMC\_SC11& OGLE050734.42-685611.4&  62456 & 0.1874672&     19.53& 19.05&  19.592& 19.219&  (1) \\ 
LMC\_SC11& OGLE050825.47-690107.6& 142863 & 0.2049147&     19.73& 19.24&  19.793& 19.411&  (1) \\ 
LMC\_SC13& OGLE050722.21-690257.0& 215876 & 0.1687843&     19.64& 19.14&  19.702& 19.311&  (1) \\ 
LMC\_SC14& OGLE050254.47-691316.9&  21599 & 0.1684096&     19.67& 19.21&  19.733& 19.379&  (1) \\ 
LMC\_SC14& OGLE050309.65-684327.6&  58832 & 0.1756374&     19.90& 19.49&  19.964& 19.659&  (1) \\ 
	 &			 &	  &	     &	        &      &	&	&      \\
LMC\_SC1 & OGLE053350.81-694413.3&  258104& 0.2928712&     19.12& 18.59&  19.178& 18.756&  (2) \\
LMC\_SC19/20& OGLE054500.13-703519.6& 171019/37093 & 0.2821258&     19.41& 18.93&  19.471& 19.098&  (2,3) \\
	 &			 &	  &	     &	        &      &	&	&      \\
LMC\_SC2 & OGLE053138.38-693446.7& 301723 & 0.1492495(0.1930783)& 19.56& 19.13&  19.622& 19.297&  (4) \\    
LMC\_SC4 & OGLE052558.13-694938.7& 180223 & 0.1612220(0.2085666)& 19.48& 19.09&  19.541& 19.255&  (4) \\ 
LMC\_SC4 & OGLE052629.98-692822.5& 348908 & 0.1551592(0.2007234)& 19.55& 19.17&  19.612& 19.335&  (4) \\ 
LMC\_SC4 & OGLE052714.00-695928.0& 393525 & 0.1601033(0.2071194)& 19.50& 19.11&  19.561& 19.275&  (4) \\ 
LMC\_SC5 & OGLE052312.05-700830.1& 120768 & 0.1580556(0.2044704)& 19.61& 19.19&  19.672& 19.357&  (4) \\ 
LMC\_SC10& OGLE051003.66-693122.6&   6076 & 0.1579053(0.2042759)& 19.44& 18.93&  19.501& 19.100&  (4) \\ 
	 &			 &	  &	    	        &      &      &	       &       &      \\
LMC\_SC3 & OGLE052751.76-695007.2&  55444 & 0.1751352(0.2814321)& 19.02& 18.63&  19.078& 18.793&  (5) \\
LMC\_SC4 & OGLE052554.16-701216.9& 133725 & 0.1627267(0.2614924)& 19.20& 18.78&  19.259& 18.943&  (5) \\
LMC\_SC6 & OGLE052023.24-692752.2&  78140 & 0.1700085(0.2731938)& 19.18& 18.84&  19.239& 19.000&  (5) \\
LMC\_SC14& OGLE050308.41-690053.5&  38054 & 0.1657659(0.2663762)& 19.26& 18.81&  19.319& 18.975&  (5) \\
  	 &			 &	  &	    	        &      &      &        &       &      \\
\hline											   
DF05     &                       &  DF05  &  DF05               & S03& S03  & DF05  &C03 &\\
         &                       &        &                     &phot. z.point& redd.&       & redd.&\\    
\hline
LMC\_A   & $-$                   &  28114 & 0.112677(0.145766)~~&   19.88&  19.41            &19.939&19.579-19.691 & (4)\\   
\hline
\end{tabular}
\label{t:table2}
\end{center}
\end{table*}
\normalsize
Notes:\\
(1) Fundamental mode pulsators.\\
(2) Double-mode pulsators. We list their longest periods (Soszy\'nski et al. 2003).\\
(3) This star appears twice in Soszy\'nski et al.'s other\_ogle.tab, namely as LMC\_SC19 star
number 171019, and LMC\_SC20 star number 37093, for its quantities we have adopted the mean
of the two values.\\
(4) First-harmonic pulsators. In brackets the fundamentalized periods assuming $P_{1}/P = $0.773.\\
(5) Stars likely pulsating in the second overtone. In brackets the fundamentalized periods 
assuming $P_{2}/P = $0.6223, $\log P_{2}/P =-0.206$ according to Jurcsik et al. (2006).\\
$\left< V_{c,o} \right>$ are intensity-averaged mean magnitudes on DF05 photometric zero-point and C03 
reddening scale (see Sect. 3.1). The two values for \#28114 correspond to two choices for the
reddening (0.116 and 0.08 mag, respectively).
\\  

\clearpage

\begin{table*}[ht]
\caption{$\delta$ Scuti stars in the LMC open cluster LW 55 (Kaluzny \& Rucinski 2003) and in the 
LMC disk (Kaluzny et al. 2006).}
\begin{center}
\footnotesize
%\tiny
\begin{tabular}{lcrlcccccc}
\hline
\hline
LMC field & Star ID& Star number & ~~~~~$P$&$\left< V \right>$&$\left< V_t \right>$&$\left< B-V \right>$&$E(B-V)$&$\left< V_{t,0} \right>$&Remarks\\
       & (1)    & (1)~~~~	 & ~~~~(2) &	              & (3) &         &   (4)  &   (5)	  &  \\  
\hline
~~LW55 & V1	& 940~~~~	 & 0.094(0.151)  &   19.98    & 19.88 &0.316       &  0.060 &   19.694  & (6)\\   
~~LW55 & V2	&1000~~~~	 & 0.098	 &   20.45    & 20.35 &0.374       &  0.118 &   19.984  & (7)\\   
~~LW55 & V3	&1044~~~~	 & 0.095	 &   20.49    & 20.39 &0.315       &  0.059 &   20.207  & (7)\\   
~~LW55 & V4	&1098~~~~	 & 0.102	 &   20.46    & 20.36 &0.328       &  0.072 &   20.137  & (7)\\   
~~LW55 & V5	&1359~~~~	 & 0.095	 &   20.27    & 20.17 &0.350       &  0.094 &   19.879  & (7)\\   
~~LW55 & V8	&1790~~~~	 & 0.092	 &   20.76    & 20.66 &0.374       &  0.118 &   20.294  & (7)\\   
       &        &                &		 &	      &       &            &        &           &     \\	 
~~Disk & V2	&1984~~~~	 & 0.0675	 &   21.15    & 21.09 &0.33	   &  0.074 &   20.861  & (7)\\    
\hline											   
\end{tabular}
\label{t:table3}
\end{center}
\end{table*}
\normalsize
Notes:\\
(1) Star Id's and numbers are from Kaluzny \& Rucinski (2003) for the LW 55 variables, and
from Kaluzny et al. (2006) for the LMC disk $\delta$ Scuti star.\\
(2) Periods for the $\delta$ Scuti stars in LW 55 were derived in the present 
paper and have errors of $\pm$ 0.02 in $\log P$, for the disk $\delta$ Scuti are from 
Kaluzny et al. (2006).\\
(3) Magnitudes corrected for the tilt of the LMC.\\ 
(4) Individual reddenings were estimated from the observed $\left< B-V \right>$ assuming
an intrinsic color $\left< B-V \right>_0$=0.256 (see Section 3).\\
(5) Tilt-corrected magnitudes, dereddened assuming $A_V$=3.1$E(B-V)$.\\
(6) Second overtone pulsator, in brackets the fundamentalized period assuming 
$P_{2}/P = $0.6223.\\
(7) Fundamental mode pulsators.\\  

\end{document}